\documentclass[12pt]{iopart}
\usepackage{iopams} 
\usepackage{graphicx}
\usepackage{pstricks} 
\usepackage{helvet}
\usepackage{url}
\usepackage{caption}

\begin{document}

\title{Binding Blocks: building the Universe one nucleus at the time}

\author{C. Aa. Diget, A. Pastore, K. Leech, T. Haylett ,  S. Lock, T. Sanders, M. Shelley, H. V. Willett,}
\address{Department of Physics, University of York, Heslington, York, Y010 5DD, United Kingdom }

\author{ J. Keegans,}
\address{E.A. Milne Centre for Astrophysics
School of Mathematics and Physical Sciences,
University of Hull,
Hull, HU6 7RX,
United Kingdom  }

\author{ L. Sinclair,}
\address{Castle Hill Hospital,
Cottingham, HU16 5JQ,
United Kingdom  }

\author{E. C. Simpson,} 
\address{Department of Nuclear Physics, Research School of Physics and Engineering, The Australian National University, Canberra ACT 2601, Australia}

\author{and the Binding Blocks collaboration \includegraphics[width=0.05\textwidth]{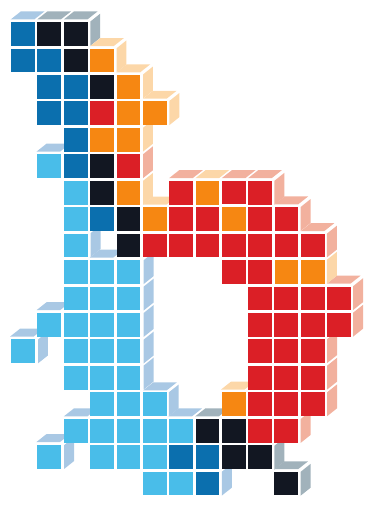}}
\ead{physics-bindingblocks@york.ac.uk}
%

\begin{abstract}
We present a new teaching and outreach activity based around the construction of a three-dimensional chart of isotopes using LEGO$^{\circledR}$ bricks\footnote{LEGO$^{\circledR}$ is a trademark of the LEGO Group of companies which does not sponsor, authorise or endorse the present work.}. The activity, \emph{Binding Blocks}, demonstrates nuclear and astrophysical processes through a seven-meter chart of all nuclear isotopes, built from over 26,000 LEGO$^{\circledR}$  bricks. It integrates A-level and GCSE curricula across areas of nuclear physics, astrophysics, and chemistry, including: nuclear decays (through the colours in the chart); nuclear binding energy (through tower heights); production of chemical elements in the cosmos; fusion processes in stars and fusion energy on Earth; as well as links to medical physics, particularly diagnostics and radiotherapy. \end{abstract}

\section{Introduction}

Nuclear physics is an exciting and interesting field, with impact across many areas of science and society, ranging from astrophysics to medicine and energy. Since the discovery of the nucleus in 1911 by Rutherford~\cite{kra12}, nuclear physicists have made enormous discoveries. The most famous is with no doubt the possibility of using nuclei as a valuable source of energy by either fusing~\cite{Edd88} or fissioning them~\cite{boh39}. These two mechanisms are not only important for industrial applications such as nuclear power plants and future fusion power, but also to explain the inner mechanism of energy production inside stars~\cite{bet67} and the production of heavy elements in the Universe~\cite{bur57}. Finally, properties of nuclei are critical for the recent achievements in nuclear medicine: from the applications of $^{99m}$Tc~\cite{wag07} to the recent developments in proton therapy~\cite{am15}.

To help communicating these important scientific achievements, we have developed a teaching and outreach activity called Binding Blocks~\cite{web2}. Taking inspiration by the original work of Ref.~\cite{kou15}, where a very detailed chart was developed as a static display, we have at University of York created an interactive activity for GCSE and A-level students and/or the public, based around the construction of a large-scale three-dimensional chart of nuclides.


As opposed to the display described in Ref.~\cite{kou15}, our activity is strongly focused on dynamic aspects of the interaction and aims at directly involving the audience to be active part in the construction process. In the present paper, we explain the main aspects of the project. The article is organised  with a definition of the nuclide chart we are going to use in Sec.~\ref{sec:nuclear}; the practical implementation of events and workshops in Sec.~\ref{sec:impl}; and a brief overview of complementary electronic material in Sec.~\ref{sec:3d}. Finally we present our conclusions in Sec.~\ref{sec:concl}.

\section{The nuclide chart}\label{sec:nuclear}

\begin{figure*}[!h]
\begin{center}
\includegraphics[width=0.7\textwidth,angle=-90]{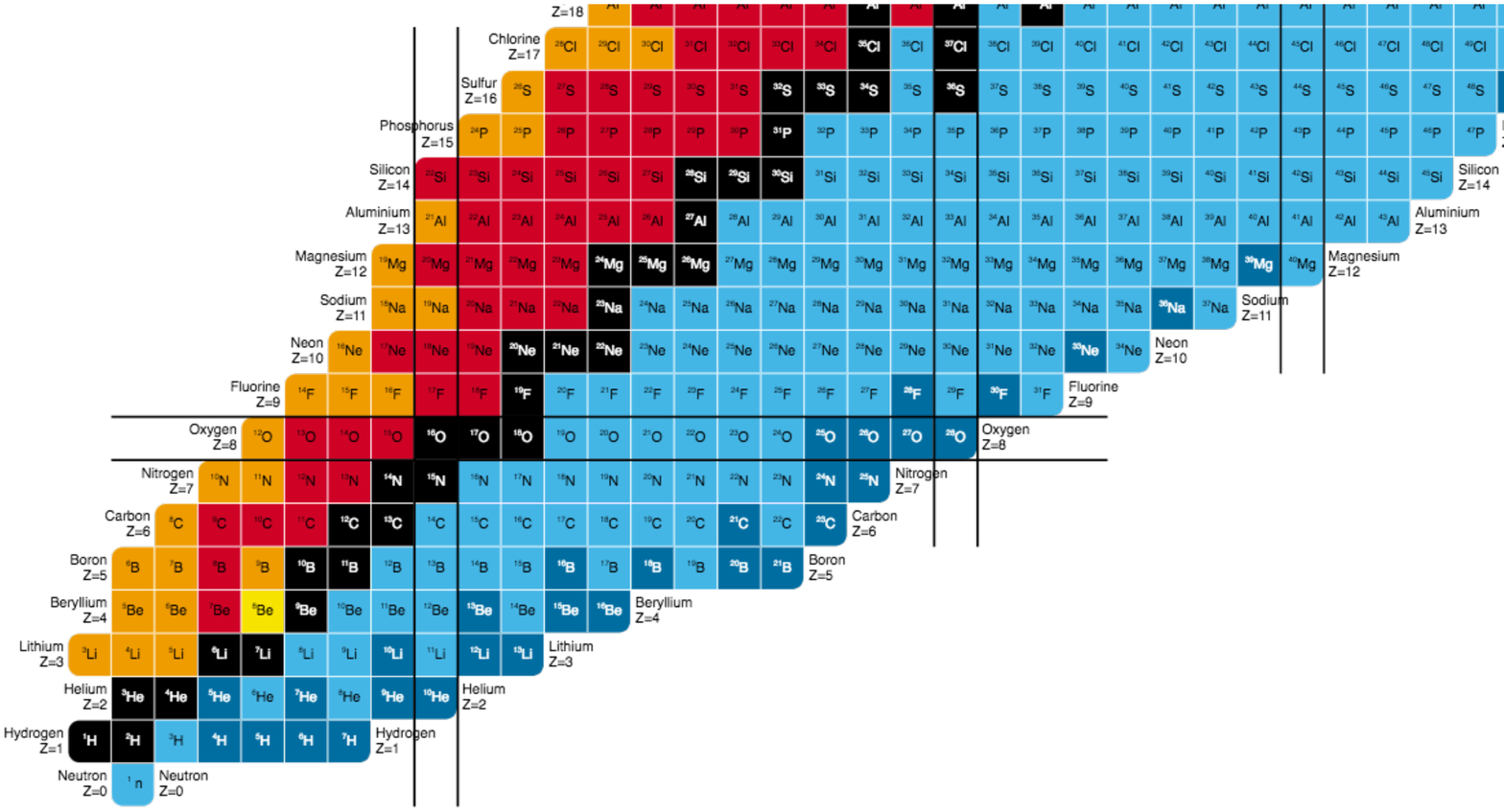}
\end{center}
\caption{(Color online) The lower part of the nuclide chart. The horizontal axis refer to the neutron number, while the vertical axis in the figure is the proton number (or element). See text for further details on data and colours. Credit to~\cite{web4}.}
\label{fig:chart}
\end{figure*}

The nuclide chart is usually defined as a two-dimensional plot where the different isotopes are ordered according to their neutron and proton number~\cite{wan12}.
In Fig.~\ref{fig:chart}, we show the low-mass part of the nuclide chart with proton number, $Z$, (chemical element) vertically and neutron number, $N$, (isotopes of each element) horizontally. The vertical and horizontal lines refer to the magic numbers: 2, 8, 20, 28,\dots for neutrons and protons (respectively)\cite{may50}. These are the nuclear equivalent of the atomic magic numbers 2, 10, 18, 36, corresponding to the noble gasses: helium, neon, argon, krypton,\dots.  

The colours in the chart refer to the different decay modes as we represent them using LEGO$^{\circledR}$ bricks. In black we represent the stable nuclei, in light blue (red) the unstable nuclei that decay via $\beta^-$ ($\beta^+$) decay~\cite{kra88}; the dark blue and orange refer to neutron emission and proton emission from unbound nuclei, usually nuclei with a very short half life. Finally, in yellow, we represent nuclei that decay via emission of an $\alpha$ particle and in green (not shown in the picture) spontaneous fission. It is worth mentioning that a nucleus in some cases can decay with different modes that compete with each other. In Binding Blocks, we have decided to make a simplification by considering only the dominant decay mode for each isotope. The data are taken from Ref.~\cite{web4}.

In Binding Blocks, we lay this 2D chart out on a table and for each isotope build towers of LEGO$^{\circledR}$ bricks of different heights, giving us a 3D nuclide chart. The height of each tower gives the excess energy available in the nucleus for nuclear decay, fission, or fusion processes. Specifically the height is chosen as the \emph{mass excess per nucleon} defined as the difference between the total mass ($M$) of the nucleus expressed in unified atomic mass units ($u$) and its mass number ($A = Z + N$). This is then divided by the total number of nucleons, giving: $(M - A\,u)/A$.

The mass excess is usually expressed in terms of the available energy, remembering Einstein's famous $E=Mc^2$, multiplying through by the speed of light squared ($c^2$) \footnote{The energy may either be given in electronVolt (eV) or Joules (J), recalling that that $1\,eV =1.602176565 \times10^{-19}\,J$}. However, noting that the mass excess is small compared to the total mass of the nucleus, and that therefore $A\,u$ is approximately the mass of the nucleus, we can express the mass excess per nucleon as available energy in Giga-Joule (GJ) per kilogram of the material, dividing by $A\,u$ in units of kilogram: $\Delta E / m = (M\,c^2 - A\,u\,c^2)/(A\,u)$. This allows us to directly compare using examples from every-day life.
In these units, each layer of LEGO$^{\circledR}$ bricks has been chosen to correspond to 25,000\,GJ/kg. In comparison, it is worth mentioning that the UK energy consumption was 125 GJ per capita in 2014 \cite{web}.

Since the mass excess is a relative measure and to avoid dealing with negative values where nuclei are stronger bound than carbon-12, which defines the atomic mass unit, all values are shifted up by adding the mass-excess per nucleon of iron-56 ($^{56}$Fe). $^{56}$Fe is the most stable isotope, with the lowest mass excess of all nuclides. This, combined with our rounding off to the nearest number of whole layers, ensures that $^{56}$Fe and the region around it is a flat region with a single layer of LEGO$^{\circledR}$ bricks. The nuclide chart then rises up towards the edges around a valley with $^{56}$Fe and the iron region at the bottom.

In our implementation, we have decided to use only experimentally observed nuclides, with the data taken from Ref.~\cite{wan12}. This represents roughly one third of existing isotopes according to most recent theoretical models~\cite{erl12}.

\section{Implementation}\label{sec:impl}

\begin{figure}
\begin{minipage}[t]{0.49\linewidth}
\vspace{0pt}
\centering
  \includegraphics[width=1.2\textwidth, angle=-90]{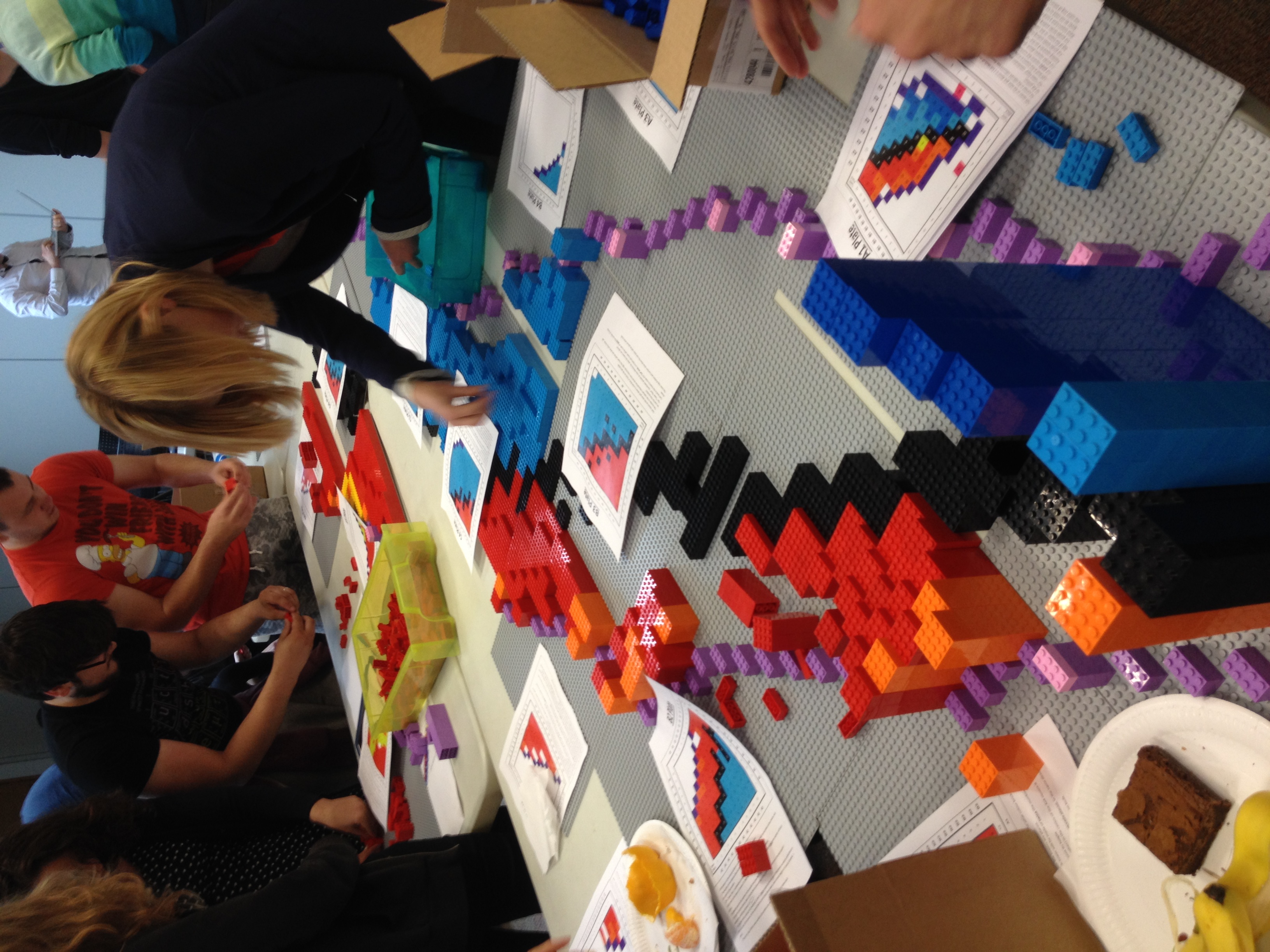}
    \caption{(Colour online) Construction of the Binding Blocks nuclide chart at a student training event (University of York, May 2016).}
\label{fig:activity}
\end{minipage}
\hspace{0.01\linewidth}
\begin{minipage}[t]{0.49\linewidth}
\vspace{0pt}
\centering
  \includegraphics[width=1.2\textwidth, angle=90]{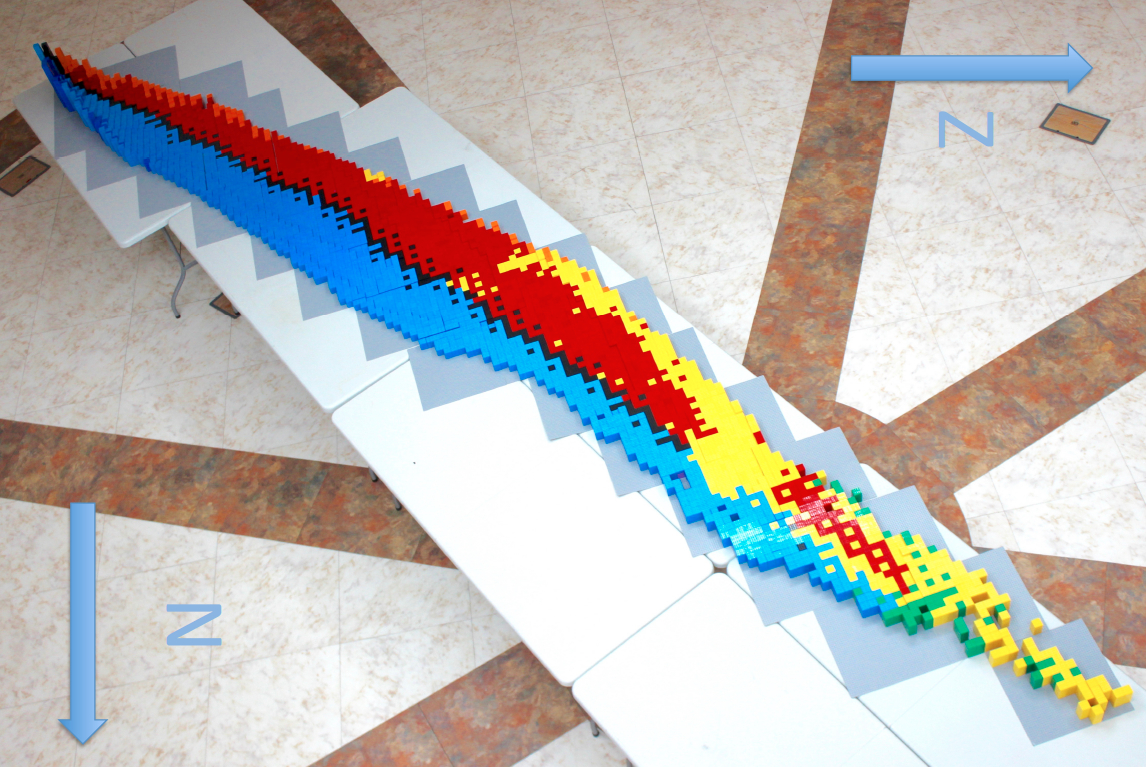}
  \caption{(Colour online) Aerial view of the nuclear chart.}
\label{fig:nslc}
\end{minipage}
\end{figure}

The key to the current project is the fact that the chart is built from scratch at each event, such that the participants actively engage in the construction and thereby get a hands-on sense of the scales involved. In Fig.~\ref{fig:activity} we show the Binding Blocks chart under construction at a student training event (University of York, May 2016), prior to their external engagement with the public using the chart. At the time of the picture, the chart is partly built up to and slightly beyond the iron region.

The full nuclide chart is shown in Fig.~\ref{fig:nslc}, as built by 50 secondary school teachers (National STEM Learning Centre, York, June 2016). The dimensions of the chart are approximately $7$\,m length and $1$\,m width.

\begin{figure*}[!h]
\begin{center}
\includegraphics[width=0.5\textwidth,angle=0]{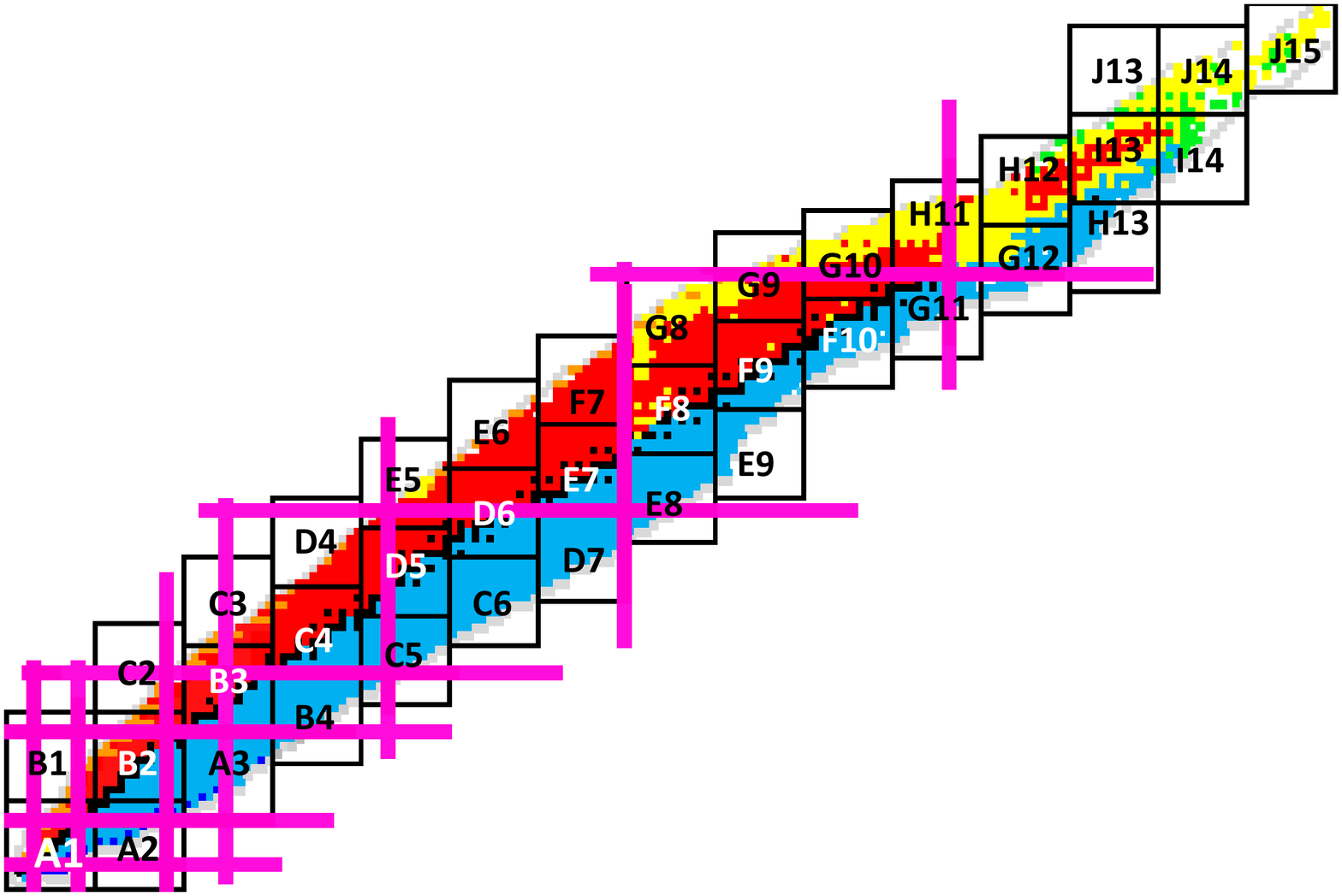}
\includegraphics[width=0.43\textwidth,angle=0]{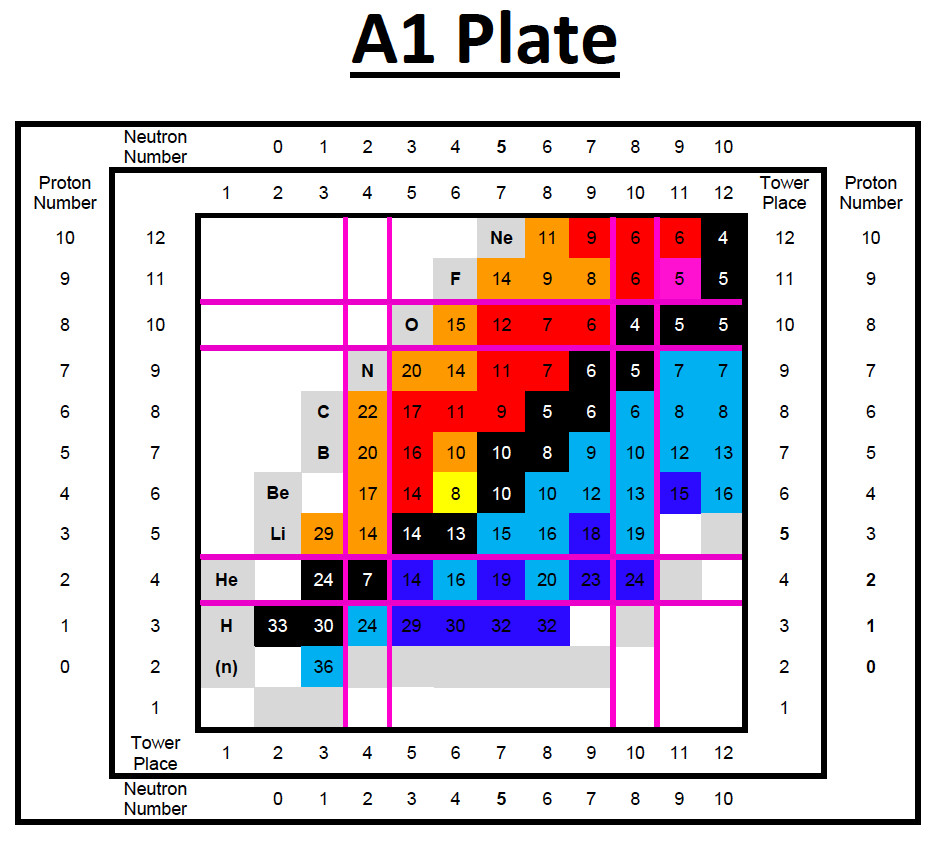}
\end{center}
\caption{(Color online) On the left panel: the instruction file used to build the nuclide chart. On the right side the detail of one of the plates. See text for details.}
\label{fig:plates}
\end{figure*}

The full chart is divided into 38 standard LEGO$^{\circledR}$ plates of approximately 38 cm $\times$ 38 cm (up to 12 by 12 towers) and each  tower is constructed from up to 36 layers of two standard two-by-four LEGO$^{\circledR}$ bricks. To construct the chart we have created an overview file shown in the left panel of Fig.~\ref{fig:plates}, as well as detailed plate layouts shown on the right. Each plate layout contains the number of neutrons and protons as well the number of LEGO$^{\circledR}$ layers for each tower. The colours on the layout correspond exactly to the colours of the bricks, as shown for the first plate on the right panel of Fig.~\ref{fig:plates}. This plate, for example, covers elements from hydrogen ($Z=1$) to neon ($Z=12$). Magic number are highlighted with coloured vertical (horizontal) lines. 

\begin{figure*}[!h]
\begin{center}
\includegraphics[width=0.5\textwidth,angle=-90]{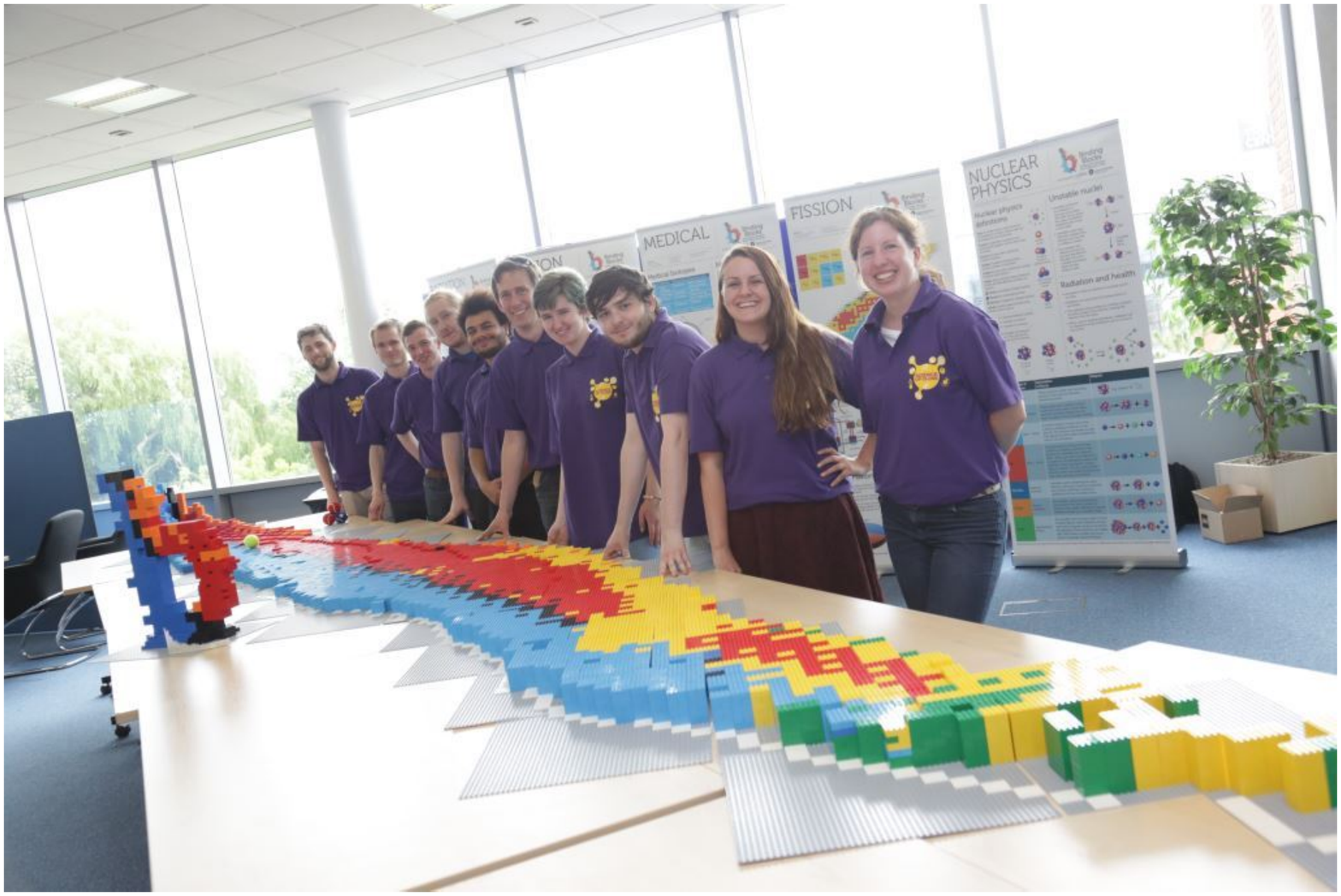}
\end{center}
\caption{Picture of the team taken at the Daresbury Laboratory Open Day (July 2016). The banners in the background are used to facilitate discussion with the participants about both physical concepts and applications.}
\label{event}
\end{figure*}

In our latest implementation we have also included the corresponding chemical symbol for isotopic chains. The plate-sheets are usually printed and given to participants in the workshop or exhibition, but they are also available electronically on tablets used during each event. This aspect is particularly important since people attending the event can navigate thought the 3D chart using the plate-sheets and identify relevant chemical elements. The construction time for the chart of nuclides during our events have ranged from 1/2 hour to 4 hours depending on the type of event. With the public at an open event, the full chart was built by participants over the course of four hours, whereas with 50 (highly competitive) teachers it was built in under one hour. A key advantage of the instructions as they stand is that the chart becomes modular. This means we can decide to build either the entire chart or build only a specific region, and similarly that if we build the full chart in an educational setting, individual groups can work on each their part of the chart. In Fig.~\ref{event}, we show the actual implementation of the chart at  STFC  Open Day at Daresbury Laboratory (July 2016)~\cite{web5}. 

The region up to and including the iron-group elements is a natural choice for smaller public events or sessions with single school classes, as it covers the bottom of the valley and thereby facilitates discussion of many astrophysical reactions, which tend to terminate in this region. Similarly, the first plate only (covering up to neon-20) already displays several isotopes of particular interest, including flourine-18 used in PET scanning \footnote{This isotope is highlighted in pink in the detailed instruction sheet, Fig.~\ref{fig:plates}, to trigger questions from participants. The decay is actually a $\beta^+$ decay, and thus it should rightly have been red, according to its decay mode.}. In all cases we would highlight the physics of light nuclei by discussing aspects related to fusion (astrophysical and on Earth), but depending on the scale of the event, other subjects can be discussed as well, up to fission which is accessible only with the full chart.

For school groups, before starting the construction, we give a short lecture of at most 10-15 minutes where we explain the main physics concepts of decay and energy and we also give instructions on how to build the chart. Normally, the workshop would be complemented by banners (see Fig.~\ref{event}) that introduce the different areas of interest across the nuclide chart, to further facilitate these discussions. These banners are available on request from our website \cite{web2}, along with other material, and they contain synthetic information on three important topics: nuclear fusion and its impact on energy production in stars and on earth; nuclear medicine and other applications, as well as nuclear fission. These topics are associated with the mass region of interest, thus according to the size of the chart we plan to build we can decide which banners to include in the event. 

The smallest set (covering up to neon-20) uses approximately 2,000 LEGO$^{\circledR}$ bricks and fills approximately a school bag. This is designed to allow a single Binding Blocks ambassador to use it with a small group of school-children and with the help of the local teacher it is possible to run such activity. The version including the iron region fills a large sports bag, uses approximately 7,000 LEGO$^{\circledR}$ bricks, and would normally be delivered by two or three ambassadors to a full school class or (with additional helpers) to a small public event. For the full chart with over 26,000 LEGO$^{\circledR}$ bricks, a car or van is necessary for transport. This version is intended for multi-school events, large-scale training events, and public events with many hundreds of visitors. As an example of this, at the recent STFC Daresbury Laboratory Open Day (July 2016), over 1,000 of the 7,500 visitors at the Open Day participated in the Binding Blocks exhibition and engaged with the construction of the chart. In this instance, the chart was completed over the course of four hours, see Fig.~\ref{event}.

\begin{figure*}[!h]
\begin{center}
\includegraphics[width=0.5\textwidth,angle=0]{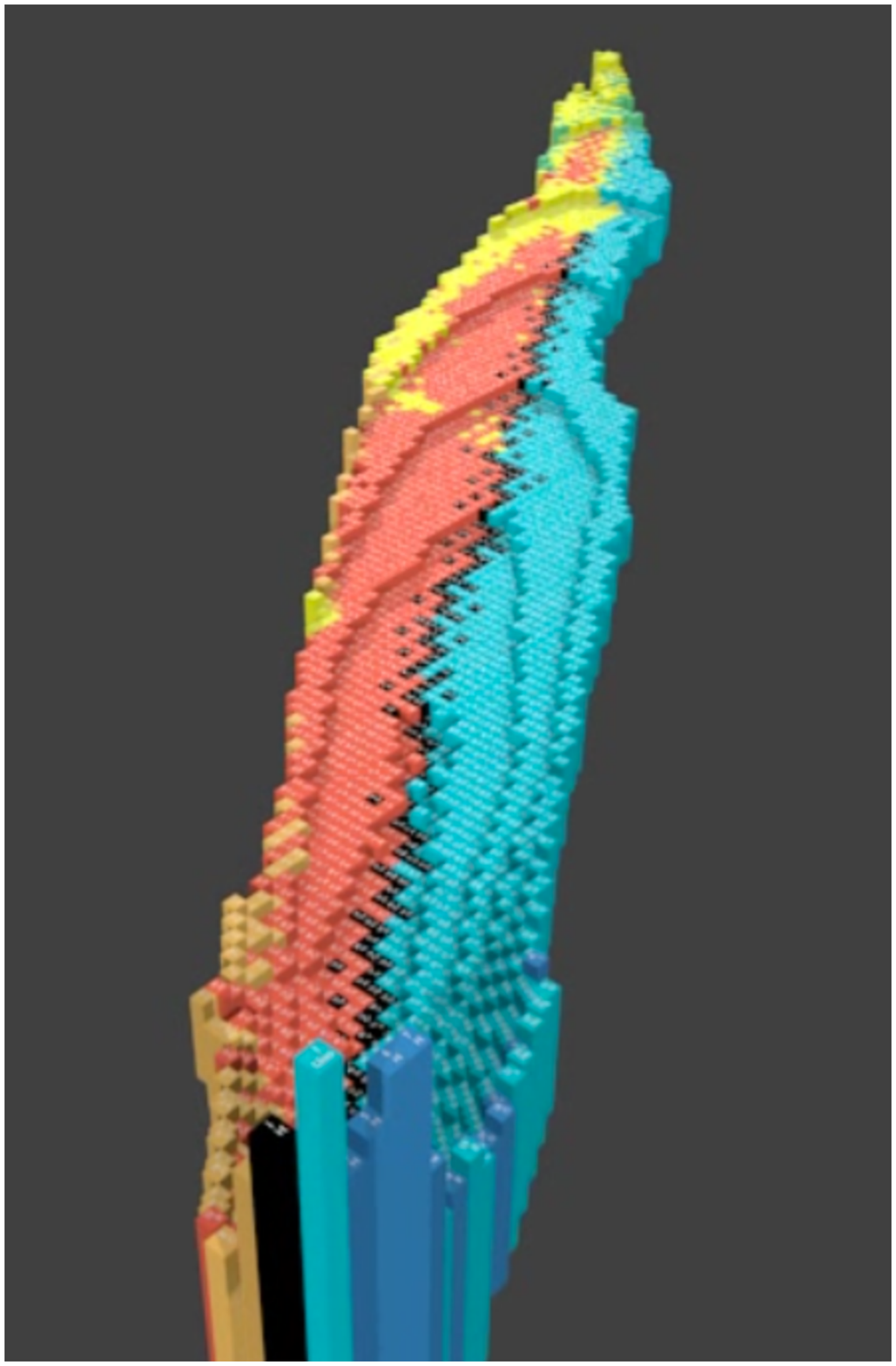}
\end{center}
\caption{ Electronic 3D version of the nuclide chart using the same colours as the real version built from LEGO$^{\circledR}$ bricks.}
\label{3d}
\end{figure*}

The cost of this activity can be easily estimated and it is essentially related to the cost of each individual brick. The individual bricks can be purchased to the online shop~\footnote{LEGO$^{\circledR}$ shop: https://shop.lego.com/en-GB/}. Given the modularity of the project it is possible to adapt the size of the chart to the available budget. The full implementation shown in Fig.~\ref{event} costs approximately $£3,500$.

\section{3D animated version}\label{sec:3d}

Using the same input data, we have furthermore created an animated 3D electronic version of the nuclide chart, by using the software Blender~\cite{web3}. The main reason for having an electronic version is the possibility of showing the full chart to participants, particularly early in the construction at large-scale events, while the chart is not still incomplete. The final result is therefore seen, and ambassadors can explain the main physical properties across the chart, also in the early stages of the construction process. 

We have thus created a short flyby movie using the 3D chart to highlight the differences of height of the different towers corresponding to individual isotopes. We kept the same colour coding and tower size of the LEGO$^{\circledR}$ brick version, but we have been able to label each tower with the name of the nuclear isotope. The resulting movie is created in mp4 format, so any generic player can visualise it and we can thus easily show it on tablets or other visual support during the exhibitions and workshops in schools. In Fig.~\ref{3d}, we show an image of the 3D chart taken from such a movie. The full movie is available as additional material on our web-page~\cite{web2}, but also on our YouTube$^{\circledR}$ channel: Binding Blocks. 

\section{Conclusions}\label{sec:concl}

In this present article, we have presented the main features of the Binding Blocks project. 
The project has been developed at the Physics Department of the University of York and it has already been delivered in several types of events: both using small scale versions or the full nuclide chart. Key material for delivery of the workshop can be downloaded from the Binding Blocks website~\cite{web2}, such that other groups will be able to profit from the material and further support is available for interested partners by contact through the Binding Blocks email address.

During each event, participants have immediately engaged with the chart, which immediately generates a wide range of questions about many aspects of nuclear physics. During training events for teachers, the participating teachers have come up with many ideas for complementing the workshop with activities in physics, chemistry, and mathematics, where the completed chart of nuclides facilitates further learning. These ideas are still under development, but include: find an isotope (or chemical element) scavenger hunts; identification of decay chains; calculations of binding energy; calculation of energy from fusion reactions; work on integrations and/or volumes (depending on the year group); as well as abundances of isotopes and/or chemical elements.

\section*{Acknowledgments}
We would like to thank E. S. Cunningham (Science \& Technology Facilities Council) and M. Langley (National STEM Learning Centre) for support in developing the teaching and outreach activity. The Binding Blocks project has been funded by EPSRC and University of York through an EPSRC Impact Acceleration Award, and by an STFC Public Engagement Small Award (ST/N005694/1).

\section*{References}
\bibliographystyle{ieeetr}
\bibliography{biblio}

\end{document}